\documentclass[aps,prd,preprintnumbers,nofootinbib,twocolumn]{revtex4}
\usepackage{bm}
\usepackage{latexsym}
\usepackage{dcolumn}
\usepackage{amsmath,amsfonts,amssymb}
\usepackage{amsthm}

\def\be {\begin{equation}}
\def\ee {\end{equation}}
\def\bea {\begin{eqnarray}}
\def\eea {\end{eqnarray}}
\def\bc {\begin{center}}
\def\ec {\end{center}}
\def\bfg {\begin{figure}}
\def\efg {\end{figure}}
\def\bi {\begin{itemize}}
\def\ei {\end{itemize}}
\def\nn {\nonumber}

\def\la {\label}
\def\le {\left}
\def\ri {\right}

\def\fr {\frac}

\def\no {\noindent}

\def\vs {\vspace}

\def\ul {\underline}

\def\a  {\alpha}
\def\b  {\beta}

\def\d  {\delta}
\def\D  {\Delta}

\def\beq{\begin{equation}}
\def\eeq{\end{equation}}
\def\br{\begin{eqnarray}}
\def\er{\end{eqnarray}}
\newcommand{\eel}[1] {\label{#1}\end{equation}}

\newcommand{\bdm}{\begin{displaymath}}
\newcommand{\edm}{\end{displaymath}}

\begin{document}
\title{A proposal for testing Quantum Gravity in the lab}

\author{Ahmed Farag Ali $^1$} \email[]{ahmed.ali@uleth.ca}
\author{Saurya Das $^1$} \email[ ]{saurya.das@uleth.ca}
\author{Elias C. Vagenas $^2$} \email[]{evagenas@academyofathens.gr}

\affiliation{$^1$~Theoretical Physics Group, Department of Physics,
University of Lethbridge, 4401 University Drive,
Lethbridge, Alberta, Canada T1K 3M4 \\}

\affiliation{$^2$~Research Center for Astronomy and Applied Mathematics,\\
Academy of Athens, \\
Soranou Efessiou 4,
GR-11527, Athens, Greece
}

\begin{abstract}
\par\noindent
Attempts to formulate a quantum theory of gravitation are
collectively known as {\it quantum gravity}. Various approaches to
quantum gravity such as string theory and loop quantum gravity, as
well as black hole physics and doubly special relativity theories
predict a minimum measurable length, or a maximum observable
momentum, and related modifications of the Heisenberg Uncertainty
Principle to a so-called generalized uncertainty principle (GUP).
We have proposed a GUP consistent with string theory, black hole
physics and  doubly special relativity theories and have showed
that this modifies all quantum mechanical Hamiltonians. When
applied to an elementary particle, it suggests that the space
that confines it must be quantized, and in fact that all
measurable lengths are quantized in units of a fundamental length
(which can be the Planck length). On the one hand, this may signal
the breakdown of the spacetime continuum picture near that scale,
and on the other hand, it can predict an upper bound on the
quantum gravity parameter in the GUP, from current observations.
Furthermore, such fundamental discreteness of space may have
observable consequences at length scales much larger than the
Planck scale. Because this influences all the quantum Hamiltonians
in an universal way, it predicts quantum gravity corrections to
various quantum phenomena. Therefore, in the present work we
compute these corrections to the Lamb shift, simple harmonic
oscillator, Landau levels, and the tunneling current in a scanning
tunneling microscope.


%
\end{abstract}

\maketitle


\section{INTRODUCTION}


An intriguing prediction of various theories of quantum gravity
(such as string theory) and black hole physics is
the existence of a minimum measurable length.
This has given rise to the so-called
generalized uncertainty principle (GUP) or equivalently,
modified commutation relations between position coordinates
and momenta. The recently proposed {\it doubly special
relativity} (DSR) theories on the other hand, also suggest a similar
modification of commutators. The commutators that are consistent
with string theory, black holes physics, DSR, {\it and}
which ensure $[x_i,x_j]=0=[p_i,p_j]$ (via the Jacobi identity)
have the following form \cite{ADV} (see Appendix)
\footnote{
The results of this article
do not depend on this particular form of GUP chosen,
and continue to hold for a large class of variants,
so long as an ${\cal O}(\a)$ term is present in the right-hand side
of Eq.(\ref{comm01}).
}
\bea
[x_i, p_j] = i \hbar\hspace{-0.5ex} \left(  \delta_{ij}\hspace{-0.5ex}
- \hspace{-0.5ex} \alpha\hspace{-0.5ex}  \le( p \delta_{ij} +
\frac{p_i p_j}{p} \ri)
+ \alpha^2 \hspace{-0.5ex}
\le( p^2 \delta_{ij}  + 3 p_{i} p_{j} \ri) \hspace{-0.5ex} \ri)
\label{comm01}
\eea
where
$\alpha = {\alpha_0}/{M_{Pl}c} = {\alpha_0 \ell_{Pl}}/{\hbar},$
$M_{Pl}=$ Planck mass, $\ell_{Pl}\approx 10^{-35}~m=$ Planck length,
and $M_{Pl} c^2=$ Planck energy $\approx 10^{19}~GeV$.
\par\noindent
In one dimension, Eq.(\ref{comm01}) gives to ${\cal O}(\a^2)$
\bea
&& \Delta x \D p \geq \frac{\hbar}{2}
\le[
1 - 2\a <p> + 4\a^2 <p^2>
\ri] ~ \nn \\
&\geq& \hspace{-1ex}
\frac{\hbar}{2} \hspace{-1ex}
\le[ 1\hspace{-0.5ex}  +\hspace{-0.5ex}  \le(\hspace{-0.5ex}  \frac{\a}{\sqrt{\langle p^2 \rangle}} +4\a^2 \hspace{-0.5ex} \ri)
\hspace{-0.5ex}  \D p^2 \hspace{-0.5ex}
+\hspace{-0.5ex}  4\a^2 \langle p \rangle^2 \hspace{-0.5ex}
- \hspace{-0.5ex}  2\a \sqrt{\langle p^2 \rangle}
\ri]\hspace{-1ex} . \label{dxdp1}
\eea
Commutators and inequalities similar to (\ref{comm01}) and (\ref{dxdp1})
were proposed and derived respectively in \cite{guppapers,kmm,kempf,brau,sm,Hossain:2010wy,cg,dv}.
These in turn imply a minimum measurable length {\it and} a maximum measurable momentum -
the latter following from the assumption that $\Delta p$ characterizes the maximum momentum of a particle
as well \cite{adler}, and also from the fact that DSR predicts such an maximum
(to the best of our knowledge, (\ref{comm01}) and (\ref{dxdp1}) are the only forms which imply both)

\bea
\D x &\geq& (\D x)_{min}  \approx \alpha_0\ell_{Pl} \la{dxmin} \\
\D p &\leq& (\D p)_{max} \approx \frac{M_{Pl}c}{\a_0}~. \la{dpmax}
\eea
\par\noindent
Next, defining (see Appendix)
\bea x_i = x_{0i}~,~~
p_i = p_{0i} \le( 1 - \a p_0 + 2\a^2 p_0^2 \ri)~, \la{mom1}
\eea
%
with $x_{0i}, p_{0j}$
satisfying the canonical commutation relations
%
$ [x_{0i}, p_{0j}] = i \hbar~\delta_{ij}, $
%
it can be shown that Eq.(\ref{comm01}) is satisfied. Here,
$p_{0i}$ can be interpreted as the momentum at low energies
(having the standard representation in position space, i.e. $p_{0i} = -i
\hbar \partial/\partial{x_{0i}}$) and $p_{i}$ as that at higher energies.\\
%
%
%
%
It is normally assumed that the dimensionless parameter
$\a_0$ is of the order of unity, in which case
the $\a$ dependent terms are important only when
energies (momenta) are comparable to the Planck energy (momentum),
and lengths are comparable to the Planck length.
However, if we do not impose this condition {\it a priori}, then this may signal the
existence of a new physical length scale of the order of
$\a\hbar=\a_o\ell_{Pl}$. Evidently, such an intermediate length
scale cannot exceed the electroweak length scale $\sim 10^{17}~\ell_{Pl}$ (as
otherwise it would have been observed) and this implies that $\alpha_{0}\leq 10^{17}$.
\par\noindent
Using Eq.(\ref{mom1}), a Hamiltonian of the form
\bea H &=& \fr{p^2}{2m} + V(\vec r)
\eea
can be written as
\bea
H&=&H_0 + H_1 + {\cal O}( \a^3) ~ \\
%
%
\mbox{where}~H_0 &=& \fr{p_0^2}{2m} + V(\vec r)   \\
\mbox{and}~ H_1 &=& -\fr{\a}{m}~p_0^3~+ \frac{5\a^2}{2m}~p_0^4~.
\eea
Thus, we see that {\it any} system with a
well-defined quantum (or even classical) Hamiltonian $H_0$ is perturbed
by $H_1$, defined above, near the Planck scale.
Such corrections extend to
relativistic systems as well \cite{Das:2010zf},
and given the robust nature of GUP,
will continue to play a role irrespective of what other quantum gravity corrections
one may consider. In other words, they are in some sense universal.

The relativistic Dirac equation is modified in a similar way and
confirms the main results of our paper \cite{Das:2010zf}.
In this paper, we first study the effects of the above GUP-corrected
Hamiltonian to a particle in a box, to ${\cal O}(\alpha)$ in Sec. IA, and
to ${\cal O}(\alpha^2)$ in Sec. IB, and show that they lead to virtually identical conclusions.
In Sec. II, we study the effects of GUP-corrected Hamiltonian to the Landau
levels. In Sec. III, we calculate the corrections due to GUP in the context of a simple harmonic oscillator.
In Sec. IV, we study the effects of GUP on the Lamb shift.
Furthermore, we compute the GUP corrections  on the tunneling current
in a scanning tunneling microscope for a step potential in Sec. V
and for a potential barrier in Sec. VI.
Finally, we summarize our results in the concluding section.


\subsection{\textbf{Solution to order $\a$}}
\par\noindent
In this subsection, we briefly review our work in \cite{ADV}.
The wave function of the particle satisfies the following
GUP-corrected Schr\"odinger equation inside the box of length $L$
(with boundaries at $x=0$ and $x=L$), where
$V(\vec r)=0$ (outside, $V=\infty$ and $\psi=0$)
\be
H \psi = E \psi~
\ee
which is now written, to order $\a$, as
\bea
&& d^2\psi + k^2 \psi + 2i\alpha \hbar d^3 \psi = 0~
\la{seqn}
%
%
%
\eea
where $d^n$ stands for $d^n/dx^n$ and $ k=\sqrt{2mE/\hbar^2}$.
A trial solution of the form $\psi=e^{mx}$ yields
\be
m^2 + k^2 + 2i\a \hbar m^3 =0~
\ee
with the following solution set to leading order in $\a$: $m=\{ik',-ik'',i/2\a\hbar\}$,
where $k'=k(1+k\a\hbar)$ and $k''=k(1-k\a\hbar)$.
Thus, the general wavefunction to leading order in $\ell_{Pl}$ and $\a$ is of the form
\bea
\psi &=& Ae^{ik'x} + B e^{-ik''x} + C e^{ix/2\a\hbar}  . \la{barrierpsi1}
\la{wavefn1}
\eea
Although the first two terms can be considered as perturbative corrections over the standard
solutions, the appearance of the new oscillatory third term is noteworthy here, with characteristic wavelength
$4\pi \a\hbar$ and momentum $1/4\alpha = M_{Pl}c/4\alpha_0$ [which is Planckian for $\alpha_0 = {\cal O}(1)$].
This can be termed a nonperturbative solution as the exponent contains $1/\alpha$ and
results in the new quantization mentioned above.
Note that however, as explained in \cite{ADV} and \cite{Das:2010zf}, $C$ scales as a power of $\alpha$,
and the
new solution disappears in the $\alpha \rightarrow 0$ limit.
\par\noindent
Imposing the appropriate boundary conditions, i.e. $\psi=0$ at $x=0$, $L$, with $A$ assumed real without loss
of generality, we get, to leading order, the following two series of solutions
($C=|C| e^{-i\theta_C}$):
\bea
\frac{L}{2\alpha\hbar} &=& \frac{L}{2\a_0 \ell_{Pl}} = n\pi + 2q \pi +2\theta_C
\equiv p\pi + 2\theta_C \label{soln1}\\
\frac{L}{2\alpha\hbar} &=& \frac{L}{2\a_0 \ell_{Pl}} = -n\pi + 2q\pi
\equiv p \pi  ~, \label{soln2}\\
p &\equiv& 2q \pm n \in \mathbb{N}. \nn
\eea
These show that there cannot be even be a single particle in the box,
unless its length is quantized as above. For other lengths, there is no
way to probe or measure the box, even if it exists. Hence, effectively
all measurable lengths are quantized in units of $\a_0\ell_{Pl}$.
We interpret this as space essentially having a discrete nature.
Note that the above conclusion holds for any unknown but {\it fixed} $\theta_C$,
which, however, determines the minimum measurable length, if any.
It is hoped that additional physically motivated or consistency conditions will
eventually allow one to either determine or at least put reasonable bounds on it.

The minimum length is $\approx \a_0 \ell_{Pl}$ in each case. Once
again, if $\a_0 \approx 1$, this fundamental unit is the Planck
length. However, current experiments do not rule out discreteness
smaller than about a thousandth of a Fermi, thus predicting the
previously mentioned bound on $\a_0$. Note that similar
quantization of length was shown in the context of loop quantum
gravity in \cite{thiemann}.
%
%
%
%
%
%
\subsection{Solution to order $\a^2$}

\par\noindent
We extend the previous solution to include the $\a^2$ term in one dimension. Working to ${\cal O}(\a^2)$,
the magnitude of the momentum at high energies as given by Eq.(\ref{mom1}) reads
\bea
p=p_0(1-\a p_0 + 2\a^2 p_0^2)~.
\label{eqn31}
\eea
The wavefunction satisfies the following GUP-corrected Schr\"odinger equation
\bea
&& d^2\psi + k^2 \psi + 2i\hbar \a d^3 \psi - 5 \hbar^2 \a^2 d^4 \psi= 0
\label{eqn32}
%
%
%
\eea
where
$ k=\sqrt{2mE/\hbar^2}$ and $d^{n}\equiv d^{n}/dx^{n}$.
%
\\
Substituting $ \psi (x)= e^{mx}$, we obtain
 \bea
 m^2 + k^2 + 2 i \a \hbar m^3 - 5 (\a\hbar)^2 m^4 =0
 \label{eqn33}
 \eea
with the following solution set to leading order in $\a^2$: $m=\{ik',-ik'',\frac{2+i}{5 \a \hbar},  \frac{-2+i}{5 \a\hbar }\}$,
where $k'=k(1+k\a\hbar)$ and $k''=k(1-k\a\hbar)$.
Thus, the most general solution to leading order in $\ell_{Pl}^2$ and $\a^2 $ is of the form
\bea
\nn \psi (x) &=& A e^{i k^{\prime} x} + B e^{-i k^{\prime\prime} x} + C e^{(2+i)x/5\a \hbar}\\&+& D e^{(-2+i)x/5\a \hbar}~.
\label{eqn34}
\eea
Note again the appearance of  new oscillatory terms, with characteristic wavelength $10\pi\a\hbar$, which
as before, by virtue of $C$ and $D$ scaling as a power of $\alpha$,
disappear in the $\alpha \rightarrow 0$ limit.
In addition, we absorb any phase of $A$ in $\psi$ so as $A$ to be real. The boundary condition
\be
\psi(0) = 0
\label{eqn35}
\ee
implies
\be
A+B+C+D=0
\label{eqn36}
\ee
and hence the general solution given in Eq.(\ref{eqn34}) becomes
\bea
\nn \psi (x)&=& 2 i A \sin(k x) e^{i \a k^2 \hbar x}-(C+D) e ^{-i k^{\prime\prime} x}
\\  &+& e^{\frac{ix}{5\a\hbar}}[C e^{\frac{2x}{5\a \hbar}} + D e^{\frac{-2x}{5\a \hbar}}].
\label{eqn37}
\eea
If we now combine Eq.(\ref{eqn37}) and the remaining boundary condition
\be
\psi(L) = 0
\label{eqn38}
\ee
we get
\bea
\nn 2 i A \sin(k L) &=& (C+ D)e^{-i[\a k^2 \hbar L+  k^{\prime\prime} L]}
\\  &-& \Big[C e^{\frac{2L}{5\a \hbar}} + D e^{\frac{-2L}{5\a \hbar}}\Big]e^{\frac{iL}{5\a \hbar}} e^{-i\a k^2 \hbar L}~.
\label{eqn39}
\eea
We can consider the exponentials $e^{-i\a k^2 \hbar L} \approx 1 $,  otherwise, since they are multiplied with $C$
or $D$, terms of higher order in $\a$ will appear. Therefore,  we have
($C = |C|e^{-i\theta_C}, D = |D|e^{-i\theta_D}$)
\bea
\nn 2 i A \sin(k L) &=& \Big[|C|e^{-i\theta_C} + |D| e^{-i\theta_D}\Big] e^{-ikL}
\\ &-&\Big[|C| e^{-i\theta_C} e^{\frac{2L}{5\a \hbar}} + |D| e^{-i\theta_D} e^{\frac{-2L}{5\a \hbar}}\Big]e^{\frac{iL}{5\a\hbar}}.~~~~~
\label{eqn40}
\eea
Now, equating the real parts of Eq.(\ref{eqn40}) (remembering that $A\in\mathbb{R}$), we have
\bea
0~~\hspace{-2ex}&=&~~\hspace{-2ex}|C| \cos(\theta_C +k L)+ |D| \cos(\theta_D + k_L) \nn\\
\hspace{-2ex}&-&~~\hspace{-2.5ex} e^{\frac{2L}{5\a\hbar}}|C| \cos(\theta_C -\frac{L}{5\a\hbar})
-e^{\frac{-2L}{5\a\hbar}} |D| \cos (\theta_D - \frac{L}{5\a\hbar}).~~~~~
\label{eqn41}
\eea
%
%
%
Note that the third term in the right hand side
dominates over the other terms in the limit $\a \rightarrow 0$.
Thus we arrive at the following equation to leading order
\be
\cos(L/5 \a \hbar - \theta_ C) = 0 ~.
\label{eqn42}
\ee
This implies the quantization of the space  by the following equation
\bea
\frac{L}{5\a \hbar} = (2p+1) \frac{\pi}{2} + \theta_C ~,\hspace{2ex} p \in &\mathbb{N}~.
\label{eqn43}
\eea
\noindent
Once again, even though the $\a^2$ term has been included, the space quantization given in Eq.(\ref{eqn43})
suggests that the dimension of the box, and hence all measurable lengths are
quantized in units of $\a_0\ell_{Pl}$, and if
$\a_0 \approx 1$, this fundamental unit is of the order of Planck length.
And as before, the yet undetermined constant $\theta_C$ determines the minimum measurable length.

\section{The LANDAU LEVELS}
\par\noindent
Consider a particle of mass $m$ and charge $e$ in a constant
magnetic field ${\vec B} = B {\hat z}$, described by the vector
potential ${\vec A}=Bx {\hat y}$ 
and the Hamiltonian
\bea H_0 &=& \frac{1}{2m}\le( \vec p_0 - e \vec A\ri)^2  \la{lanham1}\\
&=& \frac{p_{0x}^2}{2m} + \frac{p_{0y}^2}{2m} - \frac{eB}{m}~x p_{0y} +
\frac{e^2 B^2}{2m}~x^2~. \la{lanham2}
\eea
Since $p_{0y}$ commutes with $H$, replacing it with its eigenvalue
$\hbar k$, we get
\be H_0 = \frac{p_{0x}^2}{2m} + \frac{1}{2} m \omega_c^2 \le( x -
\frac{\hbar k}{m \omega_c}\ri)^2 ~\la{lanham4}\ee
where $\omega_c=eB/m$ is the cyclotron frequency. This is nothing
but the Hamiltonian of a harmonic oscillator in the $x$ direction,
with its equilibrium position given by $x_0 \equiv \hbar k/m
\omega_c$. Consequently, the eigenfunctions and eigenvalues are
given, respectively, by
\bea \psi_{k,n} (x,y) &=& e^{ik y} \phi_n (x-x_0) \\
E_n &=& \hbar \omega_c \le( n +\frac{1}{2} \ri)~,~n\in \mathbb{N}~
\eea
where $\phi_n$ are the harmonic oscillator wavefunctions.
\par\noindent
The GUP-corrected Hamiltonian assumes the form \cite{dv}

\bea
\nn H &=& \frac{1}{2m}\le( \vec p_0 - e \vec A\ri)^2 -
\frac{\a}{m}\le( \vec p_0 - e \vec A\ri)^3  \\ \nn &&+ \frac{5\a^2}{2m} \le( \vec p_0 - e \vec A\ri)^4
\la{lanham13}
\eea
\bea
&=& H_0 - ~\sqrt{8 m }~ \a~ H_0^{\frac{3}{2}}+~10~\a^2 ~m~ H_0^2
%
\eea
where in the last step we have used Eq.(\ref{lanham1}).
Evidently, the eigenfunctions remain unchanged. However, the eigenvalues are shifted by

\bea
\Delta E_{n(GUP)}&=&   \langle \phi_n |- \sqrt{8 m } ~\a H_0^{\frac{3}{2}}+6~\a^2 m H_0^2| \phi_n \rangle
= \nn \\&-&\sqrt{8 m} ~\a ~ (\hbar \omega_c )^{\frac{3}{2}}
 \left(n+\frac{1}{2}\right)^{\frac{3}{2}} \nn \\ &+& 10~ m \a^2 (\hbar \omega_c)^2 \left(n+\frac{1}{2}\right)^2 \\ \nn
\eea
\par\noindent
which can be written as
\bea
\frac{\Delta E_{n(GUP)}}{E_n^{(0)}}&=& -\sqrt{8 m} ~\a  (\hbar \omega_c )^{\frac{1}{2}}
 \left(n+\frac{1}{2}\right)^{\frac{1}{2}}\nn \\  &+&  10~ m \a^2 (\hbar \omega_c) \left(n+\frac{1}{2}\right)~.
\eea

\par\noindent
For n=1, we obtain the following relation

\bea
\frac{\Delta E_{1(GUP)}}{E_1^{(0)}} &=&  - \frac{\sqrt{12 m}~ (\hbar \omega_c )^{\frac{1}{2}}}{M_{Pl} c}~\a_0
 \nn \\&+&  \frac{15~ m~ (\hbar \omega_c)}{M_{Pl}^2 c^2}~\a_0^2  ~.
\eea
\par\noindent
For an electron in a magnetic field of $10 T$, $\omega_c \approx 10^3 GHz$

\bea
\frac{\Delta E_{1(GUP)}}{E_1^{(0)}} \approx - 10^{-26} \a_0 + 10^{-52} \a_0^2 ~.
\eea
\par\noindent
Thus, quantum gravity/GUP does affect the Landau
levels. However, once again, assuming $\a_0 \sim 1$ renders
the correction too small to be measured. Without this
assumption, due to an accuracy of one part in $10^3$ in direct measurements of Landau levels using a scanning tunnel
microscope $(STM)$ (which is somewhat optimistic) \cite{wildoer}, the
upper bound on $\a_0$ becomes

\be
\a_0 < 10^{23}~.
\la{beta2}
\ee
Note that this is more stringent than the one derived in previous
works \cite{dv}.
\section{SIMPLE HARMONIC OSCILLATOR}
\par\noindent
We now consider a particle of mass $m$. The Hamiltonian of the
simple harmonic oscillator with the GUP-corrected Hamiltonian assumes the form

\be
H= H_0 + H_1 =\frac{p_0^2}{2m} + \frac{1}{2} m \omega^2 x ^2 - \frac{\a}{m} p_0^3 + \frac{5 \a^2}{2} p_0^4 ~.
\ee
\par\noindent
Employing time-independent perturbation theory, the eigenvalues are shifted up to the first order
of $\a$ by

\be
\Delta E_{GUP} = \langle \psi_n |H_1| \psi_n \rangle
\ee
where $\psi_n$ are the eigenfunctions of the simple harmonic oscillator
and are given by

\be
\psi_n(x) = \left(\frac{1}{2^n n!}\right)^{\frac{1}{2}} \left(\frac{m \omega}{\pi \hbar}\right)^{\frac{1}{4}}
e^{-\frac{ m \omega x^2}{2 \hbar}} H_n ( \sqrt{\frac{m\omega}{\hbar}} ~x)
\ee
where
\be
H_n(x) = (-1)^n e ^{x^2} \frac{d^n}{dx^n} e^{- x^2}
\ee
are the Hermite polynomials.
\par\noindent
The $p_0^3$ term will not make any contribution to first order because it is an odd function and thus, it gives
a zero by integrating over a Gaussian integral. On the other hand, the $p_0^4$ term will
make a nonzero contribution to first order.
The contribution of the $p_0^4$ term to first order is given by
\be
\Delta E_{0(GUP)}^{(1)} = \frac{5 \a^2}{2 m} <\psi_{0}\mid \hbar^4 \frac{d^4}{dx^4}\mid\psi_{0}>
\ee
and thus we get
\be
\Delta E_{0(GUP)}^{(1)} = \frac{5 \a^2 \hbar^4}{2 m}\left(\frac{\gamma}{\pi}\right)^{\frac{1}{2}}
\hspace{-1ex}\gamma^{2}\hspace{-1ex}\int dx~e^{- \gamma x^2} (3-6 \gamma x^2 +  \gamma^2 x^4)
 \ee
where $\gamma$ is equal to $\frac{ m\omega}{\hbar}$.
\par\noindent
By integrating, we get the shift of the energy to first order of perturbation as follows
\be
\Delta E_0^{(1)} = \frac{15}{8} \hbar^2 \omega^2 m \a^2
\ee
or, equivalently,
\be
\frac{\Delta E_0^{(1)}}{E_0^{(0)}} = \frac{15}{4} \hbar \omega m \a^{2}~.\la{DE1}
\ee
\par\noindent
We now compute the contribution of the $p^3$ term to second order of perturbation
\be
\Delta E_n^{(2)} = \sum_{k \neq n} \frac{ \mid < \psi_k \mid V_1 \mid \psi_n > \mid ^2}{E_n^{(0)} - E_k^{(0)}}
\ee
where
\be
V_1 =  i\frac{\a}{m}  \hbar^3  \frac{d^3}{dx^3}~.
\ee
\par\noindent
In particular, we are interested in computing the shift in the ground state energy
to second order
\be
\Delta E_0^{(2)} = \sum_{k \neq n} \frac{ \mid < \psi_k \mid V_1 \mid \psi_0 > \mid ^2}{E_0^{(0)} - E_k^{(0)}} \la{DE}
\ee
and for this reason we employ the following properties of the harmonic oscillator eigenfunctions
\be
< \psi_m \mid  x \mid \psi_n >  =
\left \{
\begin{array}{ll}
0~, &  m \neq n \pm 1\\
\sqrt{\frac{n+1}{2 \gamma}}~, & m =n+1 \\
\sqrt{\frac{n}{2\gamma}}~,& m=n-1
\end{array}
\right. \la{mx}
\ee
and
\bea
< \psi_m \mid  x^3 \mid \psi_0 > &=& \sum_{k,l} <\psi_m \mid x \mid \psi_k><\psi_k\mid x \mid \psi_l > \nn\\
&&< \psi_l \mid x \mid \psi_0> \la{mx3}
\eea
which is nonvanishing for the $(l,k,m)$ triplets: $(1,0,1)$, $(1,2,1)$, and $(1,2,3)$.

\par\noindent
Thus, the ground state energy is shifted by
\bea
\Delta E_0^{(2)} = \frac{ \a^2 \hbar^6}{ m^2}
\sum_{m \neq 0} \frac{ \mid < \psi_m \mid \frac{d^3}{dx^3} \mid \psi_0 > \mid ^2}{E_0^{(0)} - E_m^{(0)}}~.
\eea
Since the eigenfunction $\mid \psi_0 >= \le(\frac{m \omega}{ \pi \hbar}\ri)^{1/4} e^{-\frac{m \omega}{2 \hbar} x^2}$,
we have $ \frac{d^3}{dx^3}\mid \psi_0 >=\le(3 \gamma^2 x - \gamma^3 x^{3}\ri) \mid \psi_0 > $.
By employing these into Eq.(\ref{DE}), we get

\bea
\Delta E_0^{(2)} = \frac{ \a^2 \hbar^6}{ m^2} \gamma^4
\sum_{m \neq 0} \frac{ \mid < \psi_m \mid (3 x - \gamma x^{3}) \mid \psi_0 > \mid ^2}{E_0^{(0)} - E_m^{(0)}}~.
\eea
Using Eqs.(\ref{mx}) and (\ref{mx3}), the energy shift  finally takes the form
\be
\Delta E_0^{(2)} = -\frac{11}{2} \a^2 m \left(E_0^{(0)}\right)^2
\ee
or, equivalently,
\be
\frac{\Delta E_0^{(2)}}{E_0^{(0)}} = - \frac{11}{2} \a^2 m  E_0^{(0)}= - \frac{11}{4} \a^2 m \hbar \omega ~. \la{DE2}
\ee
It is noteworthy that there are some systems that can be represented by the Harmonic oscillator such as
heavy meson systems like charmonium \cite{SHO}. The charm mass is $m_c \approx ~1.3~ GeV/c^2$  and
the binding energy $\omega$ of the system is roughly equal to the energy gap separating adjacent
levels and is given by $\hbar\omega \approx 0.3 GeV$.
The correction due to GUP can be calculated at the second order of $\a$. Using Eqs.(\ref{DE1}) and (\ref{DE2}),
we found  the shift in energy  is given  by

\bea
\frac{\Delta E_0^{(2)}}{E_0^{(0)}} = \a_0^2 \frac{m~ \hbar ~ \omega~}{ M_{Pl}^2~ c^2}
\approx  2.7~ \times~ 10^{-39}~ \a_0^2
\eea
\par\noindent
Once again, assuming $\a_0 \sim 1$ renders the correction too small to be measured.
On the other hand, if such an assumption is not made, the current accuracy
of precision measurement in the case of $J/\psi$ \cite{PDG} is at the level of $ 10^{-5}$.
This sets the upper bound on $\a_0$ to be
\bea
\a_0< 10^{17}~.
\eea
\par\noindent
It should be stressed that this bound  is
in fact consistent with that set by the electroweak scale.
Therefore, it could signal a new and intermediate length
scale between the electroweak and the Planck scale.

\section{THE LAMB SHIFT}
\par\noindent
For the Hydrogen atom, $V(\vec r) = -k/r$ ($k=e^2/4\pi
\epsilon_0=\alpha \hbar c$,
$e=$ electronic charge). 
To first order, the perturbing Hamiltonian
$H_1$, shifts the wavefunctions to \cite{bransden}
\bea
|\psi_{nlm} \rangle_1 =
|\psi_{nlm} \rangle +\hspace{-5ex}
\sum_{\{n'l'm'\}\neq\{ nlm\}}
\frac{e_{n'l'm'|nlm}}{E_{n}^{(0)} -
E_{n'}^{(0)}} |\psi_{n'l'm'}\rangle \la{wavefn1}
\eea
where $n,l,m$ have their usual significance, and
$e_{n'l'm'|nlm} \equiv \langle \psi_{n'l'm'}|H_1|\psi_{nlm}\rangle$~.

\par\noindent
Using the expression $p_0^2=2m[H_0 + k/r]$ \cite{brau}, the perturbing Hamiltonian reads
\be
 H_1 = - (\a \sqrt{8m}) \le[H_0 +\frac{k}{r}\ri] \le[H_0 + \frac{k}{r}\ri]^{\frac{1}{2}}.
\ee

\par\noindent
So for GUP effect to $\a$ order, we have

\be
e_{n'l'm'|nlm}~=~ \langle \psi_{n'l'm'}|\left( -\frac{\a}{m}\right) p_0^2 p_0^{} |\psi_{nlm}\rangle~.
\ee
\par\noindent
It follows from the orthogonality of spherical harmonics that the
above are nonvanishing if and only if $l'=l$ and $m'=m$

\be
e_{200|100}~=~ 2 i \a \hbar~~ \langle \psi_{200}|\le[H_0 +\frac{k}{r}\ri]\le( \frac{\partial}{\partial r}\ri)|\psi_{100}\rangle~.
\ee

\par\noindent
We utilize the following to calculate the shift in the energy:\\

(i) the first term in the sum in Eq.(\ref{wavefn1}) ($n'=2$)
dominates, since $E_n = -E_0/n^2~(~E_0=e^2/8\pi \epsilon_0 a_0 =
k/2a_0= 13.6~eV~,~a_0= {4\pi \epsilon_0 \hbar^2}/{me^2} = 5.3
\times
10^{-11}~\mbox{metre}$~,~$m=$ electron mass $=0.5~MeV/c^2$),

(ii) $\psi_{nlm} (\vec r) = R_{nl} (r) Y_{lm}(\theta,\phi)$,

(iii) $R_{10}=2 a_0^{-3/2}e^{-r/a_0}~,~\mbox{and}~~\\
~~~~~~~~~~R_{20}=(2a_0)^{-3/2}\le(2 - r/a_0\ri) e^{-r/2a_0}$,

(iv) $Y_{00}(\theta,\phi)=1/(\sqrt{4\pi})$~.\\


\par\noindent
Thus, we derive

\bea
e_{200|100}&=& - i \frac{2  \a \hbar k  }{ a_0}~~  \langle \psi_{200}|~\frac{1}{r}~|\psi_{100}\rangle\\
&=& - i \frac{8 \sqrt{2} \a \hbar k  }{27  a_0^{2}}~.
\eea

\par\noindent
Therefore, the first order shift in the ground state wavefunction is given
by (in the position representation)\vspace{-3ex}\\

\bea
\Delta \psi_{100}(\vec r) &\equiv& \psi_{100}^{(1)}(\vec
r)-\psi_{100}^{(0)}(\vec r) =
\frac{e_{200|100}}{E_1-E_2}\psi_{200} (\vec r)\nn\\
&=& i \frac{32 \sqrt{2} \a \hbar  k}{81 a_0^{2}~E_0~}~\psi_{200}(\vec r)\\
&=& i \frac{64 \sqrt{2}  \a \hbar }{81 a_0~}~\psi_{200}(\vec r) ~.
\eea
\par\noindent
Next, we consider the Lamb shift for the $n^{th}$
level of the hydrogen atom
\cite{bd}

\be
\Delta E_n^{(1)} = \frac{4\alpha^2}{3m^2} \le( \ln \frac{1}{\alpha} \ri) \le|
\psi_{nlm}(0) \ri|^2~.
\ee

\par\noindent
Varying $\psi_{nlm}(0)$, the additional contribution
due to GUP in proportion to its original value is given by
\be
\frac{\Delta E_{n(GUP)}^{(1)}}{\Delta E_n^{(1)}} =
2 \frac{\Delta|\psi_{nlm}(0)|}
 {\psi_{nlm}(0)}~.
\ee

\par\noindent
Thus, for the ground state, we obtain

\bea
\frac{\Delta E_{1(GUP)}^{(1)}}{\Delta E_1^{(1)}} &=&
\frac{64 \hbar~ \a_0}{81 a_0 M_{pl} c}\nn\\
&\approx&  1.2 \times 10^{-22}\hspace{-1ex}~\a_0~.
\eea
\par\noindent
The above result may be interpreted in two ways. First, if one assumes
$\a_0 \sim 1$, then it predicts a nonzero, but virtually
{\it unmeasurable} effect of  GUP and thus of quantum gravity. On the other hand, if such
an assumption is not made, the current accuracy of
precision measurement of Lamb shift of about
one part in $10^{12}$ \cite{brau,newton}, sets the following upper bound on $\a_0$:
\vspace{-3ex}\\
\be
\a_0 < 10^{10}~.
\la{beta1}
\ee
\par\noindent
It should be stressed that this bound is more stringent than the ones derived in  previous
examples \cite{dv}, and is in fact consistent with that set by the
electroweak scale. Therefore, it could signal a new and
intermediate length scale between the electroweak and the Planck scale.
\section{POTENTIAL STEP}
\par\noindent
Next, we study the one-dimensional potential step
given by
\be
V^{\prime}(x)= V^{\prime}_0 ~\theta(x)
\ee
where $\theta(x)$ is the usual step function. Assuming $E <
V^{\prime}_0$, the Schr\"odinger equation to the left and right of the
barrier are written,  respectively, as
\bea
d^2\psi_{<} + k^2 \psi_{<} + 2 i \a \hbar d^3 \psi_{<} = 0\\
d^2\psi_{>} - k_1^2 \psi_{>} + 2 i \a \hbar d^3 \psi_{>} = 0
\eea
where
$ k=\sqrt{2mE/\hbar^2}$ and $k_1=\sqrt{2m(V^{\prime}_0-E)/\hbar^2}~.$
\par\noindent
Considering solutions of the form $\psi_{<,>}=e^{mx}$, we get
\bea
m^2+k^2+2 i \a \hbar m^3=0 \\
m^2- k_1^2 + 2 i \a \hbar m^3= 0
\eea
\par\noindent
with the following solution sets to leading order in $\a$,
each consisting of three values of $m$
\bea
x<0 ~: m &=& \{ i k^{\prime}, - i k^{\prime\prime}, \frac{i}{2\a \hbar} \} \\
x \geq 0 ~: m &=& \{  k_1', -k_1^{\prime\prime},\frac{i}{2\a \hbar} \}
\eea
where
\bea
k^{\prime}&=& k(1+ k \a \hbar),~~ k^{\prime\prime}= k(1-k\a \hbar)\\
k_1^{\prime}&=& k_1 (1- i \a \hbar k_1),~~ k_1^{\prime\prime}= k_1 (1+ i \a \hbar k_1)~.
\eea
Therefore, the wavefunctions take the form
\bea
\psi_<&=& A e^{ik^{\prime} x} + B e^{- i k^{\prime \prime } x} + C e^{\frac{i x}{2\a \hbar}},~~~ x<0\\
\psi_>&=& D e^{-k_1^{\prime\prime} x} + E e^{\frac{i x}{2\a \hbar}},~~~0 \leq x
\eea
where we have omitted the left mover from $\psi_>$.
\par\noindent
Now the boundary conditions at $x = 0$ consist of three equations (instead of the usual two)
\be
d^n \psi_<|_0= d^n\psi_>|_0,~~~~~ n=0,1,2~.
\ee
This leads to the following conditions:
\bea
A+B+C&=& D+E \\
i\big(k^{\prime} A - k^{\prime \prime} B + \frac{C}{2\a\hbar}\big)&=& -k_1^{\prime\prime} D + \frac{i E}{2\a\hbar}\\
k^{\prime 2} A + k^{\prime \prime 2} B + \frac{C}{(2\a\hbar)^2}&=& \frac{E}{(2\a\hbar)^2}- k_1^{\prime\prime 2} D~.
\eea
%
%
%
Assuming $C \sim E \sim {\cal O}(\a^2) $, we have the following solutions to leading order in $\a$
\bea
\frac{B}{A}&=&\frac{i k^{\prime}+k^{\prime\prime}_1}{i k^{\prime\prime}-k^{\prime\prime}_1},\\
\frac{D}{A}&=&\frac{2 i k}{i k^{\prime\prime}-k^{\prime\prime}_1},\\
\frac{E-C}{(2\a\hbar)^{2}A}&=&\frac{k^{\prime 2}(i k^{\prime\prime}-k^{\prime\prime}_1)+k^{\prime\prime 2}(i k^{\prime}+k^{\prime\prime}_1)
+k^{\prime\prime 2}_1(2 i k)}{i k^{\prime\prime}-k^{\prime\prime}_1}.~~~~~~
\eea
\par\noindent
It can be easily shown that the GUP-corrected time-dependent
Schr\"odinger equation admits the following
modified conserved current density, charge density and conservation law, respectively,  \cite{dv}
\bea
J &=& \frac{\hbar}{2mi} \le( \psi^\star \frac{d\psi}{dx}
- \psi \frac{d\psi^\star}{dx}
\ri)\nn\\
&+&\hspace{0.5ex} \frac{\alpha \hbar^2}{m}
\le(
\frac{d^2|\psi|^2}{dx^2} - 3 \frac{d\psi}{dx} \frac{d\psi^\star}{dx}
\ri) \la{current1}~, \\
\rho &=& |\psi|^2~,~
\frac{\partial J}{\partial x} + \frac{\partial \rho}{\partial t} = 0 ~.
\eea
The conserved current is given as

\bea
\nn J= J_0 +J_1 &=& \frac{\hbar k}{m} \big(|A|^2-|B|^2\big) \\&+&
 \frac{2\a\hbar^2 k^2}{m} \big(|A|^2+|B|^2\big) +\frac{|C|^2}{\a m}~.
\eea

\par\noindent
The reflection and transmission coefficients are given by

\bea
R &=& \le|\frac{B}{A}\ri|^2 \frac{1-2 \a \hbar k}{1+2 \a \hbar k}\nn\\
&=& \le|\frac{i k^{\prime}+k^{\prime\prime}_1}{i k^{\prime\prime}-k^{\prime\prime}_1}\ri|^2\frac{1-2 \a \hbar k}{1+2 \a \hbar k}\nn\\
&=& \frac{(k^2+k^2_1)^2}{(k_1^2+ k^2)^2 (1-4 \a \hbar k )} \frac{1-2 \a \hbar k}{1+2 \a \hbar k}\nn\\
&=& 1.\\ \nn\\
T&=& \frac{-\frac{\a \hbar^2 k_1^2 }{m} |D|^2 e^{-2 k_1 x}+ \frac{\a \hbar^2 k_1^2 }{m} |D|^2 e^{-2 k_1 x} }
{\frac{\hbar k}{m} |A|^2(1+2\a \hbar k)}~~\\
&=& 0,\\\nn\\
R+T&=& 1.
\eea
At this point we should note that GUP did not affect R and T up to ${\cal O}{(\a)}$.

\section{POTENTIAL BARRIER }
%
%
%
%
\par\noindent
In this section we apply the above formalism to an STM and show that in an
optimistic scenario, the effect of the
GUP-induced term may be measurable.
In an STM, free electrons of energy $E$ (close to the Fermi energy)
from a metal tip at $x=0$, tunnel quantum mechanically to
a sample surface a small distance away at $x=a$. This gap
(across which a bias voltage may be applied) is associated with
a potential barrier of height $V^{\prime \prime}_0>E$ \cite{stroscio}.
Thus
\be
V^{\prime \prime}(x) = V^{\prime \prime}_0~\le[ \theta(x) - \theta(x-a) \ri]
\ee
where $\theta(x)$ is the usual step function.
The wave functions for the
three regions, namely, $x\leq 0$, $0\leq x \leq a$, and $x\geq a$, are $\psi_{1}$,$\psi_{2}$,
and $\psi_{3}$, respectively, and satisfy the
GUP-corrected time-independent Schr\"odinger equation
%
\bea
&& d^2\psi_{1,3} + k^2 \psi_{1,3} +2i\alpha \hbar d^3 \psi_{1,3} = 0\nn \\
&& d^2\psi_2 -k_1^2 \psi_2 + 2i \alpha \hbar d^3 \psi_2 = 0 \nn
\eea
where
$ k=\sqrt{2mE/\hbar^2}$ and $k_1=\sqrt{2m(V^{\prime \prime}_0-E)/\hbar^2}~.$\\
The solutions to the aforementioned equations to leading order in $\a$ are
\bea
\psi_1 &=& A e^{ik'x} + B e^{-ik''x} + P e^{ix/2\alpha\hbar} \\
\psi_2 &=& F e^{k_1'x} + G e^{-k_1''x} + Q e^{ix/2\alpha\hbar} \\
\psi_3 &=& C e^{ik'x} + R e^{ix/2\alpha\hbar}
\eea
where
$k'=k(1+\alpha\hbar k), k''=k(1-\alpha\hbar k),
k_1'=k_1(1-i\alpha\hbar k_1), k_1''=k_1(1+i\alpha\hbar k_1)$
and $A,B,C,F,G,P,Q,R$ are constants of integration.
In the above, we have omitted the left mover from $\psi_3$.
Note the appearance of the new oscillatory terms with characteristic
wavelengths $\sim \alpha\hbar$, due to the third order
modification of the Schr\"odinger equation.
The boundary conditions at $x=0,a$ are given by
\bea
d^n\psi_1|_{x=0} &=& d^n\psi_2|_{x=0}~~,~n=0,1,2~ \\
d^n\psi_2|_{x=a} &=& d^n\psi_{3}|_{x=a}~~,~n=0,1,2~.
\eea
If we assume that  $ P~\sim ~Q~ \sim~ R~\sim~ {\cal O}(\a^2) $, we get the following solutions

\bea
\frac{C}{A}&=& \frac{i(k^{\prime}k^{\prime\prime}_1+k^{\prime\prime}k^{\prime}_1+ k^{\prime}k^{\prime}_1+k^{\prime\prime}
k^{\prime\prime}_1)e^{-ik^{\prime}a+k^{\prime\prime}_1a}}{e^{(k^{\prime}_1+k^{\prime\prime}_1)a}
(k^{\prime}+i k^{\prime}_1)(k^{\prime\prime}+i k^{\prime\prime}_1)-(k^{\prime}-i k^{\prime\prime}_1)(k^{\prime\prime}
-i k^{\prime}_1)},\nn\\
\la{solB}\\
\frac{B}{A}&=&\frac{k^{\prime\prime}_1+ i k^{\prime}}{k^{\prime\prime}_1-i k^{\prime\prime}}
\le[e^{i k^{\prime}a- k^{\prime}_1 a} \frac{C}{A}-1\ri],\la{solB1}\\
\frac{F}{A}&=& \frac{(1+ i \frac{k^{\prime}}{k^{\prime\prime}_1})e^{ik^{\prime}a-k^{\prime}_1 a} \frac{C}{A}}
{1+\frac{k^{\prime}_1}{k^{\prime\prime}_1}},\\
\frac{G}{A}&=&\frac{(1-i\frac{k^{\prime}}{k^{\prime}_1})e^{i k^{\prime}a+k^{\prime\prime}_1 a}\frac{C}{A}}{1+
\frac{k^{\prime\prime}_1}{k^{\prime}_1}}~.
\eea

\par\noindent
From Eq.(\ref{current1}), it follows that the transmission coefficient
of the STM, given by the ratio of the right moving currents
to the right and left of the barrier, namely, $J_R$ and $J_L$, respectively, is
to ${\cal O}(\alpha)$
\be
T = \frac{J_R}{J_L}
= \le| \frac{C}{A} \ri|^2 - 2\alpha \hbar k \le| \frac{B}{A} \ri|^2~
\ee
%
%

\par\noindent
which gives using the solutions in Eqs.(\ref{solB}) and (\ref{solB1}) the following final expression
\bea
T &=& T_0 \le[1 + 2\a\hbar k(1-T_0^{-1}) \ri] \la{trans01}\\
T_0 &=& \frac{16E(V^{\prime \prime}_0-E)}{V_0^{{\prime \prime}2}} e^{-2k_1a}
\eea
where $T_0$ is the standard STM transmission coefficient.
The measured tunneling current is proportional to
$T$ (usually magnified by a factor ${\cal G}$),
and using  the following approximate (but realistic) values \cite{stroscio}
\bea
&& m = m_e = 0.5~MeV/c^2 ~,~
E\approx V^{\prime \prime}_0 = 10~eV  \nn \\ 
&& a = 10^{-10} ~m ~,~
I_0 = 10^{-9} ~A~,~
{\cal G} = 10^9 \nn
\eea
we get
\bea
&&\frac{\d I}{I_0} = \frac{\d T}{T_0} = 10^{-26} , \nn \\
&& \d {\cal I} \equiv {\cal G} \d I = 10^{-26} ~A
\eea
where we have chosen $\a_0=1$ and $T_0=10^{-3}$, also a fairly
typical value. Thus, for the GUP-induced
excess current $\d{\cal I}$ to give the difference of the
charge of just one electron,
$e\simeq10^{-19}~C$, one would have to wait for a time
\be
\tau = \frac{e}{\d{\cal I}}  = 10^{7}~s \la{time}
\ee
or, equivalently, about $4$ months, which can perhaps be argued to be not that
long. In fact, higher values of $\a_0$ and a more accurate estimate will likely reduce this time,
and conversely, current studies may already be able to put an upper bound on $\a_0$.

What is perhaps more interesting is the following relation between
the {\it apparent barrier height} $\Phi_A \equiv V^{\prime \prime}_0 - E$ and the
(logarithmic) rate of increase of current with the gap, which follows
from Eq.(\ref{trans01})
\be
\sqrt{\Phi_A} =
\frac{\hbar}{\sqrt{8m}} \le| \frac{d\ln I}{da}\ri|
-\frac{\a \hbar^2(k^2 + k_1^2)^2}{8m(kk_1)}e^{2k_1a}~.
\ee
Note the GUP-induced deviation from the usual
linear $\sqrt{\Phi_A}$ vs $|d\ln I/da|$ curve. The exponential
factor makes this particularly sensitive to changes in the
tip-sample distance $a$, and hence amenable
to observations.
Any such observed deviation may signal the existence of GUP and, thus,
in turn an underlying theory of quantum gravity.

\section{Conclusions}
\par\noindent
In this work we have investigated the consequences of quantum gravitational corrections
to various quantum phenomena such as  the Landau levels, simple harmonic oscillator, the Lamb shift, and
the tunneling current in a scanning tunneling microscope and
have found that the upper bounds on $\a_0$ to be $10^{23}$, $10^{17}$, and $10^{10}$ from the
first three respectively.
The first one gives a length scale bigger than electroweak length that is not right experimentally.
It should be stressed that the last three bounds are more stringent
than the ones derived in the previous study \cite{dv}, and might be consistent with that set by the electroweak scale.
Therefore, it could signal a new and intermediate length
scale between the electroweak and the Planck scale.
On the other side, we have found that even if  $\a_0 \sim 1$, we still might measure
quantum gravitational corrections in a scanning tunneling microscopic case as was shown in Eq.\ (\ref{time}).
This is in fact an improvement over the general conclusion of \cite{dv}, where it was shown
that quantum gravitational effects are virtually negligible if the GUP parameter $\beta_0 \sim 1$,
and appears to be a new and interesting result.
It would also be interesting to apply our formalism to other areas including cosmology,
black hole physics and Hawking radiation, selection rules in quantum mechanics,
statistical mechanical systems etc. We hope to report on these in the future.


\vspace{0.5cm}

\par\noindent {\bf Acknowledgments}\\
We thank A. Dasgupta, S. Hossenfelder, R. B. Mann and L. Smolin for interesting discussions.
This work was supported in part by the Natural
Sciences and Engineering Research Council of Canada and by the
Perimeter Institute for Theoretical Physics.



\section{APPENDIX}

\subsection{ Proof for Eq. (1) }
\par\noindent
Since black hole physics and string theory suggest a modified Heisenberg algebra
(which is consistent with GUP) quadratic in the momenta
(see e.g. Ref. [1] ) while DSR theories suggest one that is
linear in the momenta (see e.g. Ref. [2] ), we try to incorporate both of the
above, and start with the most general algebra with linear and quadratic terms
%
\bea
[x_i,p_j] &=& i\hbar (\delta_{ij}+\delta_{ij} \a_1 p+ \a_2 \frac{p_{i} p_{j}}{p} +
\beta_1 \delta_{ij}  p^2 \nn \\  &+& \beta_2 p_{i} p_{j})~.
\label{eqn1}
\eea
Assuming that the coordinates commute among themselves, as do the momenta, it follows
from the Jacobi identity that
\be
-\le[ [x_i,x_j],p_k \ri] =
 \le[ [x_j,p_k],x_i \ri] +
 \le[ [p_k,x_i],x_j \ri]  = 0~.\label{eqn11}
\ee
Employing Eq.(\ref{eqn1}) and the commutator identities,
and expanding the right hand side,  we get
(summation convention assumed)
\bea
 0 &=& [[x_j,p_k],x_i]+[[p_k,x_i],x_j]\nn \\&=&
i\hbar (
-\a_1 \d_{jk} [x_i,p] - \a_2 [x_i,p_j p_k p^{-1}]  -  \b_1 \d_{jk}[x_i, p_l p_l]
\nn \\&&  -\b_2 [x_i, p_j p_k]) - (i \leftrightarrow j)
\nn \\
\nn \\ &=& i\hbar
\le(
-\a_1 \d_{jk} [x_i,p]
-\a_2 (
[x_i,p_j] p_k p^{-1}  + p_j[x_i,p_k] p^{-1} \ri. \nn \\
&& \le. + p_j p_k [x_i,p^{-1}]
)
- \b_1 \d_{jk}
\le(
[x_i,p_l]p_l + p_l [x_i,p_l]
\ri)
\ri. \nn \\
&& \le. - \b_2 \le(
[x_i,p_j] p_k + p_j[x_i,p_k]
\ri)
\ri)
- (i \leftrightarrow j )~.
\label{eqn3}
\eea
To simplify the right hand side of Eq.(\ref{eqn3}), we now evaluate the following commutators

\vs{.3cm}
\no
\ul{{\bf (i) {$[x_i,p]$ to ${\cal O}(p)$} } }

\vs{.2cm}
\no
Note that
\bea
[x_i,p^2] &=& [x_i, p \cdot p] = [x_i,p]p + p [x_i,p] \la{xp2a} \\
&=& [x_i, p_k p_k ] = [x_i,p_k]p_k + p_k [x_i, p_k] \nn \\
&=& i\hbar \le( \d_{ik} + \a_1 p \d_{ik} + \a_2 p_i p_k p^{-1} \ri) p_k
+ i\hbar p_k \le( \d_{ik} \ri. \nn \\
&& \le. + \a_1 p \d_{ik} + \a_2 p_i p_k p^{-1} \ri)
~~(\text{to}~{\cal O}(p)~\text{using}~(\ref{eqn1})) \nn \\
&=& 2i \hbar p_i \le[ 1 + (\a_1+\a_2) p \ri] \la{xp2b}~.
\eea
Comparing (\ref{xp2a}) and (\ref{xp2b}), we get
\bea
[x_i,p ] = i\hbar \le( p_i p^{-1} + (\a_1 + \a_2 ) p_i \ri)~.
\la{xip}
\eea

\vs{.3cm}
\no
\ul{{\bf (ii) {$[x_i,p^{-1}]$ to ${\cal O}(p)$} } }

\vs{.2cm}
\no
Using
\bea
0=  [x_i,I] =[x_i, p \cdot p^{-1}] = [x_i,p] p^{-1} + p [x_i,p^{-1}]
\eea
it follows that
\bea
[x_i, p^{-1}] &=& - p_i^{-1}[x_i,p]p^{-1} \nn \\
&=& -i\hbar~p^{-1} \le(
p_i p^{-1} + (\a_1 + \a_2) p_i
\ri) p^{-1}  \nn \\
&=& -i\hbar~p_i p^{-3} \le(
1 + (\a_1+\a_2) p
\ri)~. \la{xipinv}
\eea

\vs{0.2cm}

\no
Substituting (\ref{xip}) and (\ref{xipinv}) in (\ref{eqn3}) and simplifying, we get
\bea
0&=&[[x_j,p_k],x_i]+[[p_k,x_i],x_j] \nn \\
&=&
\le(
(\a_1 - \a_2) p^{-1} \hspace{-0.5ex} + \hspace{-0.5ex} (\a_1^2 + 2\b_1 - \b_2) \ri) \Delta_{jki}
\eea
\par\noindent
where $\Delta_{jki} = p_i \d_{jk} - p_j \d_{ik}$.
\no
Thus one must have $\a_1=\a_2 \equiv -\a$ (with $\a>0$; The negative sign follows from
Ref. [3] of our paper), and $\b_2 = 2\b_1 + \a_1^2$. Since from dimensional grounds it
follows that $\b \sim \a^2$, for simplicity, we assume $\b_1=\a^2$. Hence
$\b_2=3\a^2$, and we get Eq.(1) of this paper, namely,
\bea
[x_i,p_j] = i\hbar \hspace{-0.5ex} \le(
\delta_{ij}\hspace{-0.5ex} - \hspace{-0.5ex} \a \hspace{-0.5ex} \le( p \d_{ij} + \frac{p_i p_j}{p} \ri)\hspace{-0.5ex}
+ \hspace{-0.5ex} \a^2 \hspace{-0.2ex} (p^2 \d_{ij} + 3 p_i p_j \hspace{-0.5ex} )
\ri)\hspace{-0.2ex}.~~~
\la{xipj}
\eea
\subsection{Proof for Eq. (5)}
\par\noindent
We would like to express the momentum $p_j$ in terms of the
{\it low energy momentum} $p_{0j}$ (such that $[x_i,p_{0j}]=i\hbar \d_{ij}$).
Since Eq.(\ref{xipj}) is quadratic in $p_j$, the latter can at most be
a cubic function of the $p_{0i}$. We start with the most general form
consistent with the index structure
\bea
p_j = p_{0j} + a p_0 p_{0j} + b p_0^2 p_{0j}~,
\label{eqn22}
\eea
where $a \sim \a$ and $b\sim a^2$.
From Eq.(\ref{eqn22}) it follows that
\bea
[x_i, p_j] &=& [ x_i , p_{0j} + a p_0 p_{0j} + b p_0^2 p_{0j}]\nn\\
&=& i\hbar \d_{ij} + a\le(
[x_i,p_0]p_{0j} + p_0 [x_i, p_{0j}]
\ri)
\nn \\&+& \hspace{-1ex} b \hspace{-0.5ex} \le(
[x_i,p_0] p_0 p_{0j} \hspace{-0.3ex} + \hspace{-0.3ex} p_0[x_i,p_0] p_{0j} \hspace{-0.3ex} + \hspace{-0.3ex} p_0^2[x_i,p_{0j}]
\ri).~~~~~~~
\label{eqn23}
\eea

\no
Next, we use the following four results to ${\cal O}(a)$ and $[x_i,p_{0j}]=i\hbar$ in Eq.(\ref{eqn23}):

\vs{.3cm}
\no
(i) $[x_i,p_0] = i \hbar~p_{0i} p_0^{-1}$, which follows from
Eq.(\ref{xip}) when $\a_i =0$, as well from the corresponding Poisson bracket.

\vs{.3cm}
\no
(ii) $p_j = p_{0j}(1+ap_0) + {\cal O}(a^2) \simeq p_{0j} (1 + a p)$ [from Eq.(\ref{eqn22})].
Therefore, $p_{0j} \simeq \frac{p_j}{1+ap} \simeq (1-ap) p_j $~.

\vs{.3cm}
\no
(iii) $p_0 = (p_{0j} p_{0j})^{\frac{1}{2}}
= \le((1-ap)^2 p_j p_j \ri)^{1/2}
= (1-a p) p~.$

\vs{.3cm}
\no
(iv) $p_{0i} p_0^{-1} p_{0j} = (1-ap) p_i (1-ap)^{-1} p^{-1} (1-ap) p_j = (1-ap) p_i p_j p^{-1}$~.

\vs{.4cm}
\no
Thus, Eq.\ (\ref{eqn11}) yields
\bea
[x_i,p_j] &=& i\hbar \d_{ij} + i a\hbar \le( p \d_{ij} + p_i p_j p^{-1} \ri)
\nn \\ &+& i \hbar (2b-a^2) p_i p_j + i \hbar (b-a^2) p^2 \d_{ij}~.
\eea
Comparing with Eq.(\ref{xipj}), it follows that $a=-\a$ and $b=2\a^2$.
In other words
\bea
p_j = p_{0j} - \a p_0 p _{0j} + 2 \a^2 p_0^2 p_{0j}
= p_{0j}\le( 1 - \a p_0 + 2 \a^2 p_0^2 \ri)~~~
\label{eqn30}
\eea
which is Eq.(5) in this paper.
%
%
%
%
%


\end{document}